\documentstyle{lamuphys}
\def\et{{\it et al.}}

\def\Mmpc{$h^{-1}$~{\rm Mpc}\qquad}
\def\kms{km s$^{-1}$}
\input epsf
\makeatletter
\let\chapter\hid@chapter
\makeatother
\begin{document}
\pagenumbering{arabic}
\title{Void Hierarchy -- a guiding principle to the study of faint 
structures in voids}

\author{Ulrich\,Lindner\inst{1}, Jaan\,Einasto\inst{2}, 
Maret\,Einasto\inst{2}, and
Klaus J.\,Fricke\inst{1}}

\institute{Universit\"ats-Sternwarte, Geismarlandstra{\ss}e 11,
D-37083 G\"ottingen, Germany
\and
Tartu Astrophysical Observatory, 
EE-2444 T\~{o}ravere, Estonia}

\maketitle

\begin{abstract}
We introduce Void Hierarchy as an important property of the 
Large--Scale Structure in the Universe and demonstrate how
it can be used to interpret observations. Moreover the void
hierarchy constraints any realistic galaxy and structure 
formation scenario.
\end{abstract}

\section{Hierarchy of voids}
Voids were defined as low density regions or, alternatively, as 
regions completely devoid of a certain type of object. 
Mean void diameters listed in Table~1 demonstrate the 
dependence of the void size on the type of object used in the 
(second) void definition. Both
definitions imply that voids are not completely empty. Thus, 
the question is meaningful whether the distribution of galaxies in 
voids is homogeneous or reveals any structure. For example,
it was concluded that Blue Compact Galaxies (BCG) from
the Second Byurakan Survey (SBS)
or other peculiar galaxies occur 
isolated within voids (Pustil'nik {\it et al.} 1995).
Such questions are very relevant concerning scenarios of
large scale structure and galaxy formation, but they are
not conclusively answered up to now.

\vskip 0.3truecm
\noindent{\bf Table~1} \  
{\small Mean diameters of voids surrounded by different 
types of object}
\vskip -0.4truecm
{{$$\vbox
{\tabskip=0.05truecm
\halign to \hsize { \hfil# & \hfil# &  \hfill# & \hfil#\cr
\noalign {\smallskip}
\noalign{\hrule}
\noalign{\smallskip}
type of object \qquad & \qquad mean void diameter  \cr
\noalign{\smallskip}
\noalign{\hrule}
\noalign {\medskip}
 rich clusters (Abell/ACO--Catalogue)     & 100 \Mmpc \cr
 poor clusters (Zwicky--Catalogue)         &  37 \Mmpc \cr
\qquad bright ($M \le -20.3$) elliptical galaxies &  30 \Mmpc \cr
       galaxies brighter than $M = -20.3$  &  23 \Mmpc \cr
       galaxies brighter than $M = -19.7$  &  16 \Mmpc \cr
       galaxies brighter than $M = -18.8$  &  13 \Mmpc \cr
\noalign{\medskip}
\noalign{\hrule}
}}$$}}
\vskip -0.3truecm

Using the second void definition we have 
studied the properties of voids surrounded by galaxies 
from three different luminosity 
(absolute magnitude $M$) limited samples. 
Three void catalogues have been compiled. Comparisons
of voids from different catalogues revealed 
that voids form a hierarchical system (cf. Lindner 
{\it et al.} 1995, A\&A 301, 329) as it is visualized in Fig.~1a). 
In this hierarchical concept apparently isolated galaxies 
in voids may have faint close neighbors which are not 
detected because of selection effects as it is shown in Fig.~1b). 

\begin{figure}
\epsfysize=8.5cm
\vskip -2.7truecm
{\epsffile{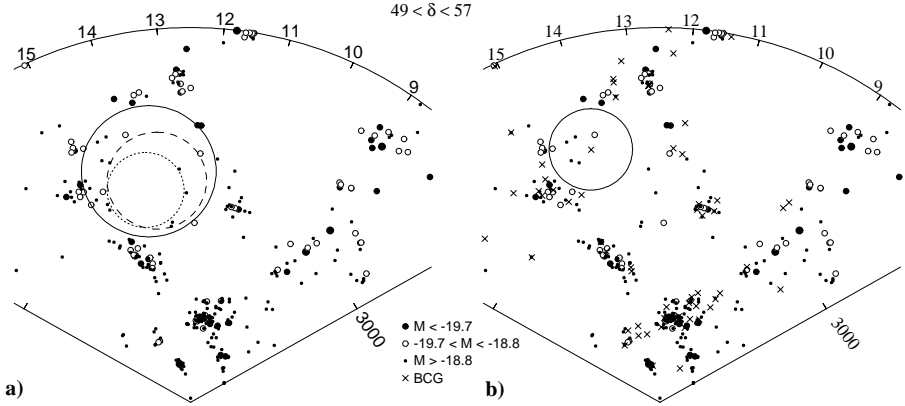}}
\vskip -0.5truecm
\caption{Wedge diagrams of a slice of the Universe 
6000 \kms\ deep and bordered by about $9^h < \alpha < 15^h$ 
and $49^\circ < \delta < 57^\circ$. 
{\bf a)} The three circles indicate an example of hierarchically interlaced 
voids defined by galaxies of different luminosity limit.
{\bf b)} Additionally BCGs from SBS are shown (crosses).
The circle indicates the distance to the nearest bright 
($M < -19.7$) neigboring galaxy.}
\end{figure}

\section{Conclusions}
By now the concept of void hierarchy is established only
for galaxies brighter than $M = -18.8$ in the nearby Universe
(up to distance $60 h^{-1}$Mpc). The study of the radial
distribution of fainter galaxies in voids along with
nearest neighbor tests (Lindner \et\ 1996) suggests
that this hierarchy continues to fainter magnitudes
and therefore contradicts a homogeneous distribution
of dwarf ga\-laxies in voids claimed by some theories 
of galaxy formation (e.g. Dekel \& Silk 1986).
With second generation instruments attached to 
the VLT (e.g. VIRMOS) it will be possible to confirm the hierarchy
of voids towards fainter luminosity limits and for more
distant regions of the Universe. The void hierarchy
itself will be helpful to devise new concepts for the
study of the large scale structure in the Universe.
%
%


\begin{thebibliography}
\bibitem{}{}{} Dekel, A., Silk, J., 1986, ApJ 303, 39
\bibitem{}{}{} Lindner, U., Einasto, J., Einasto, M., Freudling, W., 
Fricke, K.J., Tago, E., 1995, A\&A 301, 329
\bibitem{}{}{} Lindner, U., Einasto, M., Einasto, J., Freudling, W.,
Fricke, K.J., Lipovetsky, V.A., Pustil'nik, S.A.,
Izotov, Y., Richter, G.M., 1996, A\&A in press
\bibitem{}{}{} Pustil'nik, S.A., Ugryomov A.V., Lipovetsky, V.A.,
Thuan, T.X., and Guseva, N.G., 1995, ApJ 443, 499
\end{thebibliography}
\end{document}